\documentclass{optica-article}

\journal{opticajournal} 

\articletype{Research Article}

\usepackage{lineno}

\usepackage{amsmath}
\usepackage{mathtools}
\usepackage{xcolor}
\usepackage{booktabs}
\usepackage{siunitx}
\usepackage{physics}
\usepackage{subcaption}
\usepackage{geometry}
\usepackage[utf8]{inputenc}
\usepackage{float}
\usepackage{comment}
\usepackage{bm}
\usepackage{soul}

\newcommand{\CP}[1]{\textcolor{black}{#1}} 
\newcommand{\AF}[1]{\textcolor{green}{#1}} 
\newcommand{\note}[1]{\textcolor{black}{#1}} 

\begin{document}

\title{Tailoring light to create invariant modal spectra through complex channels}

\author{Cade Peters,\authormark{1,*} Isaac Nape,\authormark{1} and Andrew Forbes\authormark{1}}

\address{\authormark{1} School of Physics, University of the Witwatersrand, Private Bag 3, Wits 2050, South Africa}

\email{\authormark{*}caderibeiropeters@gmail.com} 


\begin{abstract*} 
Light’s spatial degree of freedom is emerging as a potential resource for a myriad of applications, in both classical and quantum domains, including secure communication, sensing and imaging. However, it has been repeatedly shown that a complex medium \CP{that can include the atmosphere, optical fibre, turbid media, etc.,} can perturb the spatial amplitude, phase and polarization of the structured light fields, leading to a degradation in their performance.  A promising solution to this is the use of invariant modes to which the medium appears transparent. While the creation and robustness of these modes has been experimentally demonstrated, they are difficult to implement in many important applications due to large channel matrices, a susceptibility to numerical artefacts, non-physical solutions and unreliable performance. In this work, \CP{we leverage the modal description of paraxial light beams and} outline a procedure for determining \CP{a set of modes each with an invariant modal spectrum}. \CP{Such modes can be used in place of the eigenmodes of a channel, are free form the numerical restrictions that plague calculating true eigenmodes}, are consistently realisable and require a much smaller channel matrix to compute. Using atmospheric turbulence and LG modes as the underlying basis as an illustrative example, we find robust modes for various turbulence strengths with a basis of only 231 modes, one order of magnitude smaller than previous approaches. These modes consistently show a fidelity of above 80\% after propagating through the complex channel, a significant improvement over sending the individual LG/OAM modes themselves, and reveal an invariant modal spectrum through the channel. Our approach will work for any complex medium and modal basis, paving the way for the effective implementation of the eigenmode approach in real-world situations.
\end{abstract*}

\section{Introduction}
\noindent Tailoring light's spatial degrees of freedom has led to numerous advances, allowing us to push the limits of what is possible with structured light \cite{forbes2021structured}, enabling applications such as imaging \cite{nothlawala2025quantum, angelo2019review, liu2017super}, particle tracking \cite{rosales2014measuring, ren2022compound}, classical communication \cite{wang2012terabit, bozinovic2013terabit, willner2015optical} and novel quantum encyption schemes \cite{wang2020satellite, cozzolino2019orbital, bouchard2018quantum, forbes2024quantum}. In communications, techniques such as space-division \cite{li2014space} and mode-division \cite{berdague1982mode} multiplexing have enabled the rapid increase in capacity and bandwidth of optical communication channels, whether they are terrestrial free space links \cite{krenn2014communication, trichili2020roadmap, trichili2019communicating} or through optical fibre \cite{trichili2019communicating, richardson2013space, bozinovic2013terabit}. This has provided a promising way to bridge the digital divide \cite{lavery2018sustainable} and meet the ever-increasing demands for the transfer of information in our modern age \cite{richardson2010filling}. It has also been shown that structured light modes can be used for image decomposition and reconstruction \cite{ma2021high}, opening up new avenues for faster and more accurate image recognition. It has also helped increase the resolution and sensitivity in microscopy \cite{ritsch2017orbital, furhapter2005spiral} and has been leveraged to allow for imaging past the diffraction limit of a system \cite{hell1994breaking}. 

However, the implementation of structured light is limited by several deleterious effects when passing through complex media.  For instance, organelles and fluids in biological systems limit the depth of imaging \cite{popoff2010image} and resolution \cite{vellekoop2010exploiting} in living matter, while atmospheric turbulence limits the bandwidth and reach of both classical and quantum free space optical links \cite{cox2020structured}. Much work has therefore gone into studying the impact of a complex medium on various forms of structured light, including the Bessel-Gaussian \cite{mphuthi2018bessel, lukin2014mean, bao2009propagation, watkins2020experimental }, Laguerre-Gaussian \cite{doster2016laguerre, trichili2016optical, zhao2015capacity}, Hermite-Gaussian \cite{cox2019hglg, Restuccia2016, ndagano2017comparing}. Ince-gaussian \cite{Gu2020Phenomenology} beams, but with limited success. An alternative approach is to correct the perturbation using adaptive optics \cite{zhao2012aberration, ren2014adaptive,he2022vectorial}, which can even be done in real-time without measurement using non-linear optics \cite{singh2024light,zhou2025automatic}, as well as the use of deep-learning models \cite{liu2019deep} and iterative routines \cite{li2017gerchberg}.  In addition, invariances can be exploited in vectorial light \cite{nape2021revealing}, for example, as a robust form of light for optical communication \cite{singh2023robust}, even demonstrated over real-world links \cite{peters2023invariance} as well as exploiting topologies of light \cite{shen2024optical,ornelas2024non}.  

A more general approach is to tailor the structured light field \CP{to improve its resilience through the medium.} Seeing through complex media is a highly topical field of research \cite{cao2022shaping,lib2022quantum}, and traditionally used for imaging \cite{pai2021scattering} and energy transport \cite{mosk2012controlling} by learning the transmission matrix to undo it \cite{rotter2017light,boniface2017transmission}, for example, singular value decomposition, which in the context of optical communication means that the modal basis sent and received is typically not the same \cite{bachmann2023highly}. An exciting approach is to find the true eigenmodes of the medium, which remain intact from sender to receiver \cite{shapiro1974normal}, recently experimentally demonstrated \cite{buono2022eigenmodes,klug2023robust}. All the aforementioned approaches have the problem formulated in the position basis, which comes at the expense of quasi-orthogonality, severely limiting their practical implementation in classical and quantum communication protocols as well as in imaging, where a modal basis has shown to be superior \cite{paur2016achieving,nothlawala2025quantum}.

Here we propose a new method for finding the robust forms of structured light through any complex channel by expressing the problem in a basis of choice, thereby ensuring both \CP{reliable robustness and orthogonality}. Rather than pixels, we use \CP{structured light modes} to describe modes, i.e., we determine the complex coefficients for our basis choice, which when superimposed, returns \CP{a mode whose modal content remains unchanged through the channel}. We find that the fidelity of these modes is consistently above 80\% for a variety of turbulent channels using a channel operator probed with an order of magnitude fewer modes than previous approaches. A consequence of the modal basis is the invariance of the modal spectrum to the channel, offering an alternative information encoding approach. Additionally, none of the modes exhibit the common issues associated with the pixel basis, thus making them notably more suitable for use in applications such as imaging and communications.

\begin{figure*}[htbp!]
    \centering
    \includegraphics[width=\linewidth]{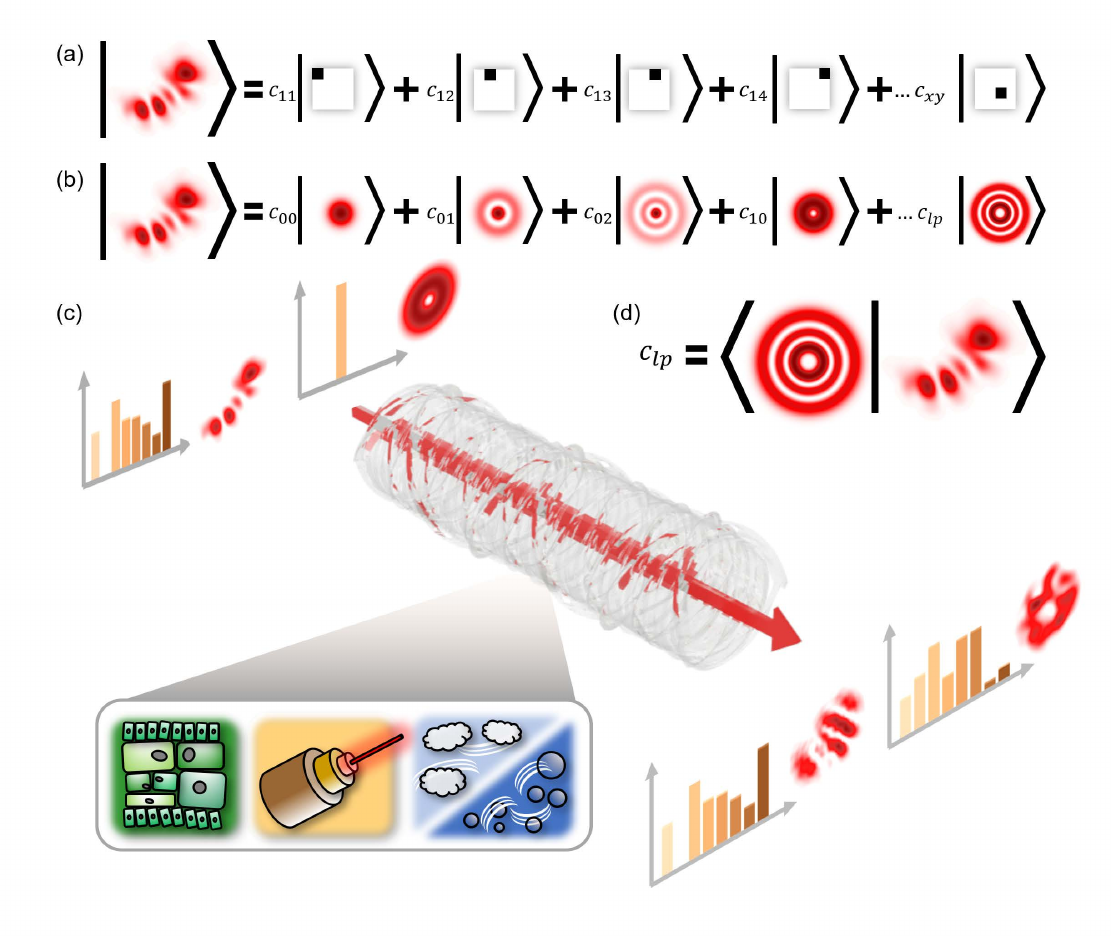}
    \caption[Eigenmodes in a modal basis]{\textbf{Eigenmodes in a modal basis.} Any complex field can be completely described by any basis with complete and orthogonal elements. Such bases include the (a) pixel/position basis and (b) a spatial mode basis such as the Laguerre-Gaussian modes.  (c) When defining an eigenmode in a spatial mode basis,  the amplitude and phase of the beam may change when propagating through the complex channel, but the modal spectrum remains unchanged. This is in contrast to a typical form of structured light, such as an LG beam, whose spatial profile and modal spectrum will change while propagating through the channel. (d) The modal spectrum is made of discrete complex coefficients that can be determined by calculating the overlap of the eigenmode with the various basis modes.} 
    \label{fig:Concept}
\end{figure*}

\section{Concept}
The traditional approach used to describe paraxial optical fields is shown in Figure \ref{fig:Concept} (a). Here, one breaks up the transverse plane into infinitely many points when using a position basis or a finite number of segments when using a pixel basis and assigns each of these a complex value, which informs the amplitude and phase of the optical field at those points. Such an approach is beneficial, as many of the governing laws and equations are written regarding position with derivatives in $x$ and $y$. However, there exist alternative ways of specifying the same field. An example is to use a basis that consists of spatial modes as shown in Figure \ref{fig:Concept} (b). In this approach, one breaks up the optical field into a linear superposition of spatial modes, where each mode is assigned a complex weighting coefficient which dictates its contribution to the total field, as seen in Figure \ref{fig:Concept} (d). Individually, each of these modes can be regarded as an eigenmode of free space,  \CP{or equivalently vacuum}. While the modal basis description of paraxial light has many advantages, e.g., for fast propagation of light \cite{sroor2021modal}, the replacement of ``space'' with ``spatial modes'' is only recently receiving attention.

\CP{In this work, we aim to combine the idea of position/pixel eigenmodes with the spatial mode description of paraxial light beams. We will refer to the set of coefficients in the linear superposition of LG modes that make up a complex field as the field's modal spectrum. This modal spectrum does not change in propagation through a uniform medium as long as one appropriately adjusts the basis modes. In complex channels where a beam might experience some perturbation, the aberration changes the modal spectrum of the field and, in so doing, perturbs its spatial profile at the output of the channel. We will therefore aim to use the eigenmode operator approach to find a modal spectrum that can propagate through a complex channel and exit completely unchanged. In this way, we can transport the full field information of the beam through the aberrating channel while minimising the loss of information. The measurement of the modal spectrum at the channel's output will allow one to reconstruct the beam at the input accurately, provided one knows the distance propagated by the beams through the channel. It is important to note that the phase and amplitude structure of these modes will change in propagation due to the natural buildup of intramodal phases between the LG basis modes, but the beam's modal content does not change. These modes can then serve the same purpose as the position/pixel eigenmodes for communicating or imaging through complex channels, but require significantly less time to calculate, are less prone to numerical artefacts, and are always physically realisable. }

\section{Theory}

\subsection{Propagation through complex media}

\CP{This section will focus on the theoretical framework necessary to accurately model a complex channel in order to verify the effectiveness of our approach numerically. It is important to note that the practical implementation of our approach in real-world channels does not require one to model the channel, but rather only to obtain the transmission matrix of the channel. This is achievable using simple projective measurements and does not need in-depth modelling of a channel's dynamics to be effectively implemented.} 

\begin{figure*}[htbp!]
    \centering
    \includegraphics[width=\linewidth]{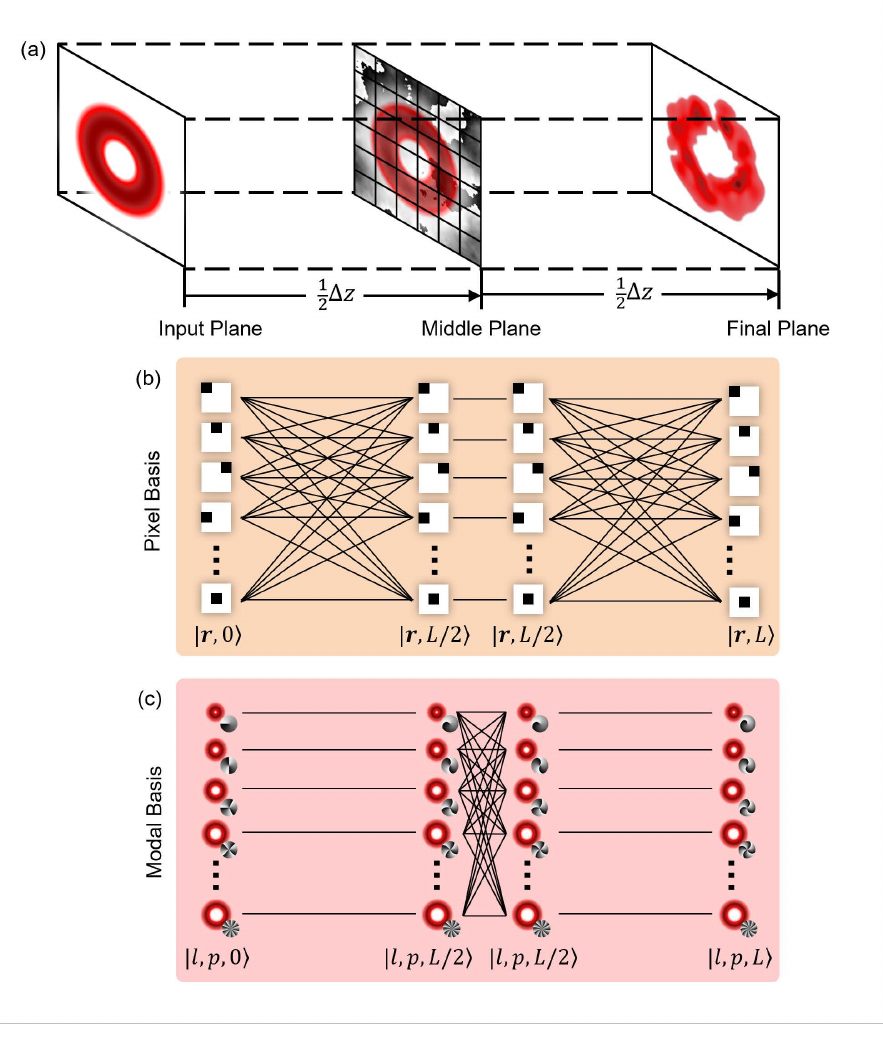}
    \caption[Operators in different bases]{\textbf{Operators in different bases.} (a) The action of a complex channel on a complex light field can be broken up into three steps: free space propagation halfway through the channel, phase aberration in the form of a single phase screen and then free space propagation through the second half of the channel.  (b) In the pixel basis this equates to each basis pixel mode in the input plane mapping to every pixel basis mode at the middle plane during free space propagation, weighted by the free space Greens function, followed by a one-to-one mapping caused by the phase aberration. This is finally followed by a second round of free space propagation which maps every pixel basis mode in the middle plane to every pixel basis mode at the output plane.  (c) In the LG basis, this is flipped around where free space propagation performs a one-to-one mapping of the spatial modes from the input plane to the middle plane. The phase aberration then causes each LG basis mode to map to every other LG basis weighted by the overlap coefficients. This is followed by a second one-to-one mapping from the middle plane to the output plane cause by free space propagation.} 
    \label{fig:Operator}
\end{figure*}

The propagation of light through a complex medium (i.e., a medium with a spatially dependent refractive index that may be changing) can be described by the stochastic Helmholtz equation,
\begin{eqnarray}
    2ik \frac{\partial }{\partial z}U(\bm{r},z) = \nabla_{\perp}^2 U(\bm{r},z) + 2k^2 \delta n (\bm{r},z) U (\bm{r},z) \,,
    \label{eq:paraxial helmholtz}
\end{eqnarray}
where $U(\bm{r},z)$ is the complex light field, $\bm{r}$ is the spatial coordinate in the transverse plane, $z$ is the propagation distance, $k$ is the wavenumber and we have assumed that the refractive index fluctuations are small, i.e., $\delta n(\bm{r},z)  = \left[n(\bm{r},z) -1 \right] \ll 1$. This equation has the same form as that of the 2D time-dependant Schr\"{o}dinger equation,
\begin{eqnarray}
    i\hbar \frac{\partial }{\partial t} \Psi(\bm{r},t) = \frac{-\hbar}{2m}\nabla_{\perp}^2 \Psi(\bm{r},t) + V(\bm{r},t) \Psi(\bm{r},t)  \,,
    \label{eq:2DTDSE}
\end{eqnarray}
 where $\Psi(\bm{r},t)$ is a 2-D wavefunction evolving in time $t$ under the influence of a  potential described by $V(\bm{r},t)$. In Equation \ref{eq:paraxial helmholtz}, $U(\bm{r},z)$ is a 2-D complex light field that evolves in propagation $z$ and is perturbed by an ``optical potential'' described by $2k^2 \delta n (\bm{r},z)$. This analogous mathematical structure allows us to apply some of the tools developed for Quantum Mechanics to the problem of paraxial light propagation. This includes the concept of time-evolving operators $e^{-iHt/\hbar}$, which, for this application, will be analogous to propagation-evolving operators. From this analogy and direct comparison between Equations \ref{eq:paraxial helmholtz} and \ref{eq:2DTDSE}, we see that the Hamiltonian for a 2-D complex light field propagating through a complex media is $H = \nabla_{\perp}^2 + 2k^2 \delta n (\mathbf{r})$. We relate the remaining factors as follows: $ \hbar \rightarrow 2k$ and $t \rightarrow \Delta z$. The propagation evolution operator thus has the form,
\begin{eqnarray}
    \hat{T}(\Delta z) &=& e^{-i\left( \Delta z \nabla_{\perp}^2 + 2k^2 \int_{z_1}^{z_2} \delta n \text{d}z \right)/2k  } = e^{-i \frac{\Delta z}{2k} \nabla_{\perp}^2 -i\Theta (\bm{r})   }, \\    
    \text{where} \quad U(\bm{r},z_2) &=& \hat{T}(\Delta z)U(\bm{r},z_1) \,,
    \label{eq:prop evolution operator}
\end{eqnarray}
where we have used,
\begin{equation}
    \Theta(\bm{\rho}) = k \int_{z_1}^{z_2} \delta n(\bm{r} , z) \text{d}z\,.
    \label{eq:Rytov_approx_phase}
\end{equation}
We therefore observe that propagation through a channel of non-uniform refractive index is the combination of two phenomena: the accumulated, spatially varying phase retardance induced by the medium's phase perturbations and mathematically described by the operator $2k\Theta (\bm{r})$, and vacuum propagation mathematically described by the operator $\nabla_{\perp}^2 $. We cannot immediately assume that both of these operators commute. However, as long as we can express the refractive index fluctuations $\delta n$ as a Taylor series in $z$ (i.e., it is an analytic function), then they can be treated as commutable with only a small loss in numerical accuracy \cite{fleck1976time}. Equation \ref{eq:prop evolution operator} can then be well-approximated with the use of a symmetrized split operator \cite{lukin2002adaptive},
\begin{eqnarray}
    \hat{T}(\Delta z) = e^{-i \frac{\Delta z}{2k} \nabla_{\perp}^2 -i\Theta (\bm{r})   } &\approx&  e^{-i \frac{\Delta z}{4k}\nabla_{\perp}^2}  e^{-i\Theta (\bm{r}) } e^{-i \frac{\Delta z}{4k}\nabla_{\perp}^2} \nonumber \\ &=& \hat{P}\left(\frac{1}{2}\Delta z\right) \hat{R}(z_1,z_2) \hat{P}\left(\frac{1}{2}\Delta z\right)\,.
    \label{eq:Symmetrized operator}
\end{eqnarray}
These results indicate that the action of our channel can be broken up into three sequential operations. The first is vacuum propagation $\hat{P}\left(\Delta z\right)$ by some distance $\frac{1}{2}\Delta z$. The second is the application of the accumulated phase perturbation $\hat{R}(z_1,z_2)$ from the entirety of the channel. The third is again vacuum propagation by a distance $\frac{1}{2}\Delta z$. This is illustrated in Figure \ref{fig:Operator} (a) for a channel of length $\Delta z$. To numerically or experimentally simulate the channel, we must begin with the initial, desired complex 2D light field and propagate it a distance $\frac{1}{2}\Delta z$. The accumulated perturbation phase is then applied in the form of a phase screen. It is then finally propagated the rest of the channel length $\frac{1}{2}\Delta z$.

It is important to note that the above approximation holds only for channels whose perturbation is small \CP{where $\sigma^2_R<1$. The exact values for the medium's correlation length $r_0$ will depend on several factors including the length of the channel and the wavelength of the light being used.} A modified approach is required when one wishes to simulate a channel with very strong perturbations and is detailed in Ref. \cite{peters2025structured}. The strong channel must be broken up into smaller segments/unit cells, where the strength of the perturbation in each unit cell would be considered to be in the weak regime ( $\sigma^2_R<1$). One then constructs an operator for each unit cell as described above. The effect of the channel is then the sequential and accumulated effect of the operators from each unit cell \cite{klug2023robust}. \CP{Our approach can easily be extended to simulating strong channels by simply calculating the operators for each unit cell, combining all of these to form the channel operator and then applying the procedures detailed below to find modes with an invariant modal spectrum. Our approach works for any channel strength in a real-world implementation as long as one is able to accurately retrieve the transmission matrix of the channel.}

\subsection{Operators in the position basis}
\CP{The propagation dynamics of an optical beam in a complex channel can be broken down into operators representing vacuum propagation and single plane refraction. In order to make use of these operators, we must now express them in a particular basis. One of the most common and familiar bases to use is the position basis and is the basis that has been previously used to determine eigenmodes.} In this basis, the initial and final complex light fields are described by some functions $U(x,y,0)$ and $U(x,y,L)$ respectively. These functions assign a complex value to every point $(x,y)$ in the initial ($z=0$) and final ($z=L$) transverse planes. The points $(x,y)$ act as our basis states as they are all orthogonal and we can fully describe any arbitrary, 2-D complex light field $\ket{U(z)}$ as a weighted, linear superposition of these points $\ket{\bm{r}}$ as,
\begin{equation}
    \ket{U(z)} = \iint U(x,y,z) \ket{\bm{r},z} \text{d}^2r \,.
\end{equation}
There then exists a propagation operator $\hat{P}$, which maps us from our complex light field in the initial plane $\ket{U(0)}$ to the complex light field in the final plane $\ket{U(L)}$. The form of this is given by,
\begin{equation}
    \hat{P}(z) = \iint g(\mathbf{r},\mathbf{r}';z) \ket{\mathbf{r}'} \bra{\mathbf{r}} \text{d}^2r\text{d}^2r' \,,
\end{equation}
which can easily be inferred from the Huygens-Fresnel diffraction integral and where $g(\mathbf{r},\mathbf{r}';z)$ is the paraxial free space Green's function given by,
\begin{equation}
    g(\mathbf{r},\mathbf{r}';z) = \frac{1}{i\lambda z} e^{\frac{i \pi}{\lambda z} |\mathbf{r} - \mathbf{r}'|} \,.
\end{equation}
The action of the operator $\hat{P}$ is to take an input state, which is a point in the input plane $\mathbf{r}$, and map its plane wave contributions to all of the output states which are all points in the output plane $\mathbf{r}'$. An output state is therefore a sum over all the plane wave contributions from each input state. 

The action of the phase aberration operator in the position basis is significantly more straightforward. Typically, one generates or has information about the spatially varying phase fluctuations of the medium $\Theta(\bm{r})$. Then, on the position basis, this phase function assigns a phase value to every point in space, representing the amount of phase gained by the input field at that point. To express this as an operator, we define it as a mapping which maps every point in space to itself, multiplied by some phase value determined by $\Theta(\bm{r})$. Mathematically, we may write the phase aberration operator as,
\begin{equation}
    \hat{R} = \int \ket{\mathbf{r}} \bra{\mathbf{r}} e^{i \Theta(\mathbf{r})} \text{d}^2r \,,
    \label{eq:postion phase}
\end{equation}
such that for a given input state $\ket{\mathbf{\rho}}$,
\begin{eqnarray}
    \hat{R}\ket{\mathbf{\rho}} = e^{i \Theta(\mathbf{\rho})} \ket{\mathbf{\rho}} \,.
\end{eqnarray}
\CP{provided that $\braket{r}{\rho} = \delta (\mathbf{r} - \mathbf{\rho})$}.

\begin{figure*}[htbp!]
    \centering
    \includegraphics[width=\linewidth]{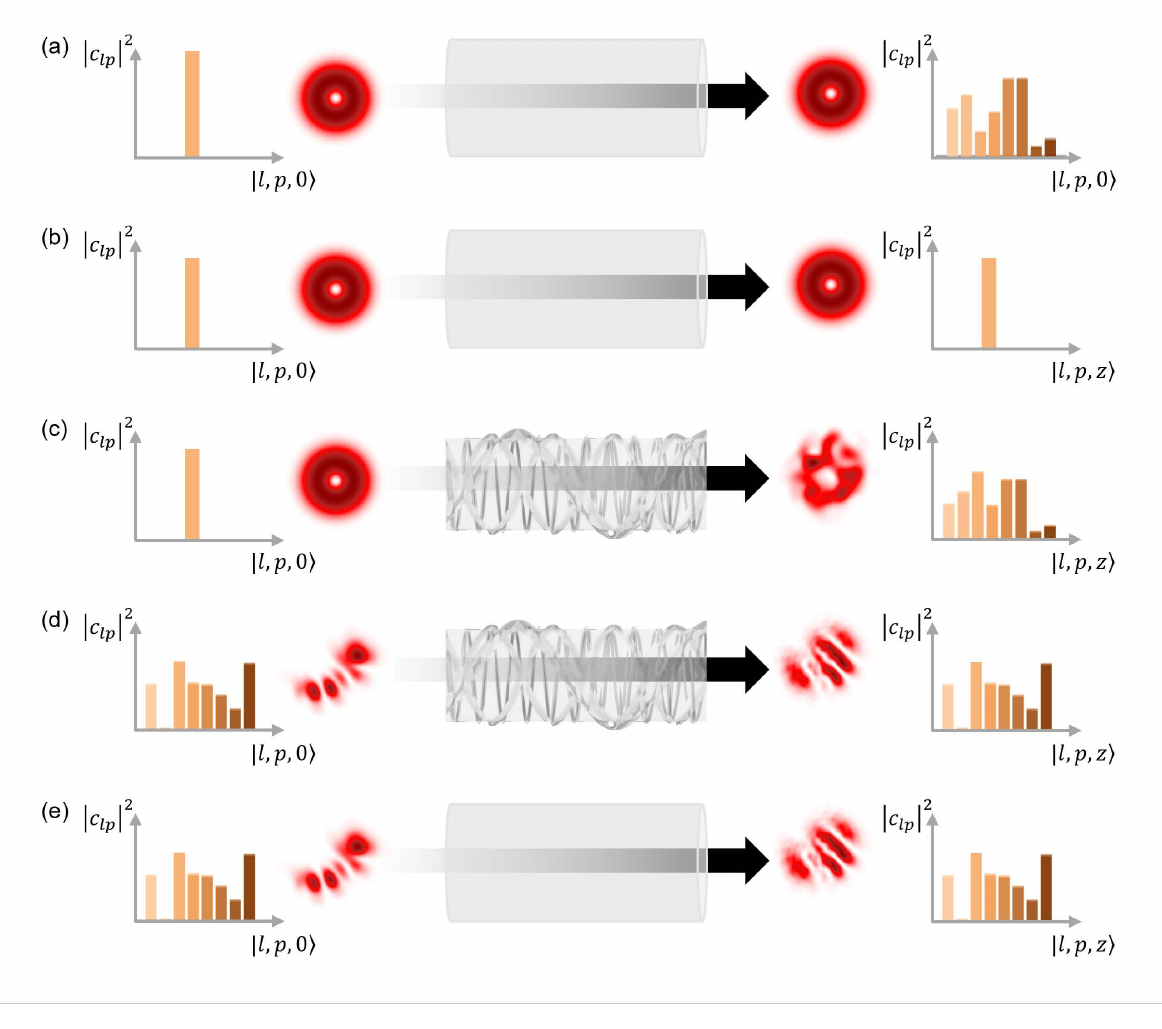}
    \caption[Propagation using spatial modes]{\textbf{Propagation using spatial modes.}\CP{(a) If one measures the modal content of a propagated LG beam in the basis it was created, then one will observe that the modal content has spread to other basis modes. (b) However, if one measures the propagated LG beam in a propagated basis, one will see that the modal content has not changed. (c) If the LG beam propagates through a complex channel, one will observe modal spreading even if one measures in a propagated basis. (d) In this work, we show that it is possible to generate a superposition of LG modes with a particular modal spectrum that, when propagated through a complex channel, will maintain the same modal spectrum at the output if measured in the propagated basis. (e) An interesting consequence of this is that these superposition modes will have the same spatial profile and modal spectrum at the output of the channel whether or not the channel has an aberration or is perfectly uniform. }} 
    \label{fig:ModalProp}
\end{figure*}

\subsection{Phase aberrations in the modal basis} \label{sec:Modal Phase}
\CP{There are several spatial modal bases that one may choose to use to express the phase aberration operator of a channel. We will choose to make use of the Laguerre-Gaussian (LG) mode family as they are a complete and orthogonal set of spatial modes whose cylindrical symmetry match the symmetries of most optical transmitters and receivers. They are also solutions to the paraxial Helmholtz equation, lending themselves well to the quantum inspired treatment proposed in this work. The LG modes are given analytically by,}
\begin{equation}
    LG_p^l(\mathbf{r},z)={\sqrt{\frac{2p!}{w^2(z)\pi(p+|l|)!}}}
           {\left(\frac{r\sqrt{2}}{w(z)}\right)^{|l|}}
           {L_p^{|l|}\left(\frac{2r^2}{w^2(z)}\right)}
           {e^{-\frac{r^2}{w^2(z)}}}
           {e^{ik\frac{r^2}{2R(z)}}}
           {e^{il\phi}}
           {e^{i(2p+|l|+1)\xi(z)}}\,,
           \label{eq:LGdef}
\end{equation}
\CP{where $p$ and $l$ are the discrete radial and azimuthal indices respectively, $L_p^{|l|}$ is the associated Laguerre polynomial, with its argument being $2r^2 / w^2(z)$, $R(z)$ is the radius of curvature and $\xi(z) = (2p+|l|+1)\arctan \left( z/z_R\right)$ is the Gouy phase which depends on the mode order $M^2 = 2p+|l|+1$. We may represent each LG beam as a basis state $\ket{l,p,z}$ uniquely indexed by $l$ and $p$ and which evolves in propagation in $z$. To find the phase aberration operator in the modal basis, we will make use of the following completeness relation,}
\begin{equation}
    \sum_{l,p} \ket{l,p,z} \bra{l,p,z} = \mathbf{I} \,,
    \label{eq:completeness}
\end{equation}
\CP{It is important to note that this completeness relationship is only valid between basis modes at the same plane.} We then substitute this into our position basis definition of the phase aberration operator (Equation \ref{eq:postion phase}) to find
\begin{eqnarray}
      \hat{R} &=& \int \ket{\mathbf{r}} \bra{\mathbf{r}} e^{i \Theta(\mathbf{r})} \text{d}^2r, \nonumber \\
      &=& \int \mathbf{I}\ket{\mathbf{r}} \bra{\mathbf{r}}\mathbf{I} e^{i \Theta(\mathbf{r})} \text{d}^2r, \nonumber \\
      &=& \int \left( \sum_{l_1,p_1} \ket{l_1,p_1,z} \bra{l_1,p_1,z} \right) \ket{\mathbf{r}} \bra{\mathbf{r}} \left(\sum_{l_2,p_2} \ket{l_2,p_2,z} \bra{l_2.p_2,z} \right) e^{i \Theta(\mathbf{r})} \text{d}^2r, \nonumber \\
      &=& \int \sum_{l_1,p_1} \sum_{l_2,p_2} \ket{l_1,p_1,z} \bra{l_1,p_1,z} \ket{\mathbf{r}}  \bra{\mathbf{r}} \ket{l_2,p_2,z} \bra{l_2,p_2,z} e^{i \Theta(\mathbf{r})} \text{d}^2r.
      \label{eq:OpInnerProducts}
\end{eqnarray}
Since the inner products $\bra{l_1,p_1,z} \ket{\mathbf{r}}$ and $\bra{\mathbf{r}} \ket{l_2,p_2,z}$ are the projections of the basis states into the position basis at the plane $z$, they may be expressed as their respective Laguerre Gaussian (LG) functions,
\begin{eqnarray}
    \braket{l_1,p_1,z}{\mathbf{r}} &=& LG^{p_1*}_{l_1}(\mathbf{r},z) \,, \label{eq:ProjRel1} \\
    \braket{\mathbf{r}}{l_2,p_2,z} &=& LG^{p_2}_{l_2}(\mathbf{r},z) \,, \label{eq:ProjRel2}
\end{eqnarray}
where $^{*}$ denotes the complex conjugate.  Using these relations, Equation \ref{eq:OpInnerProducts} becomes,
\begin{eqnarray}
    \hat{R} &=& \sum_{l_1,p_1} \sum_{l_2,p_2}  \int   LG^{p_1*}_{l_1,p_1}(\mathbf{r},z)  LG^{p_2}_{l_2}(\mathbf{r},z)   e^{i \Theta(\mathbf{r})} \text{d}^2r \ket{l_1,p_1,z}  \bra{l_2,p_2,z} \\ &=& \sum_{l_1,p_1} \sum_{l_2,p_2}  a^{l_1,p_1}_{l_2,p_2} \ket{l_1,p_1,z}  \bra{l_2,p_2,z}  \,. \label{eq:ModalPhaseAbbOp}
\end{eqnarray}
Where we define the coefficient,
\begin{equation}
    a^{l_1,p_1}_{l_2,p_2} = \int   LG^{p_1*}_{l_1}(\mathbf{r},z)  LG^{p_2}_{l_2}(\mathbf{r},z)   e^{i \Theta(\mathbf{r})} \text{d}^2x. \label{eq:ModalPhaseAbbCoeff}
\end{equation}
We interpret this operator as taking in an input basis state $\ket{l_2,p_2,z}$, aberrating it with the phase aberration given by $e^{i \Theta(\mathbf{r})}$ and then checking its overlap with the basis states at the output $\ket{l_1,p_1,z}$. The coefficient $a^{l_1,p_1}_{l_2,p_2}$ gives the degree to which an input basis mode is aberrated at the output.  This is illustrated in Figure \ref{fig:Operator} (c), where free space propagation is a direct one-to-one mapping from the input to the output plane, but the phase aberration is where each basis mode is mapped to each other basis mode, where the weightings of these mappings are given by the coefficient $a^{l_1,p_1}_{l_2,p_2}$. It is interesting to note that the action of the channel on the pixel basis complements the action of the channel in our new formulation. For the pixel basis the one-to-one mapping occurs due to the phase aberration while in the modal basis, it occurs during propagation. Critically, the mapping that mixes the basis modes in the pixel basis occurs during propagation, while in the modal basis it occurs due the phase aberration. \CP{These are the mappings that are most susceptible to errors and that have much stronger restrictions to ensure the accuracy of the mapping. For propagation, a pixel at the input plane will map to a large number of pixels in the output plane whereas for a phase aberrations, the spatial mode will typically map only to several adjacent modes in the spectrum \cite{klug2021orbital}. Thus, by using the spatial modes, we are formulating the transmission matrix in a basis that more simply describes the total action/mapping of the channel subsequently simplifying the computation and reducing the number of modes needed to accurately capture the dynamics of the channel.}

Because the process of adding the phase aberration is far less prone to numerical errors and artefacts, and has far fewer restrictions for forming a valid operator, we expect that the operator described in the modal basis will be significantly more accurate and have a much simpler description of the channel than the operator in the pixel basis. We then subsequently require a smaller channel matrix and expect eigenmodes that perform more reliably and are free from many of the inaccuracies introduced by the numerical artefacts.

\subsection{Modal Propagation} \label{sec:Modal Propagation}

\CP{An alternative to the Fresnel and angular spectrum propagation techniques used to model optical field propagation is modal propagation. This approach decomposes a 2D optical field into a sum of structured light modes, propagates these basis modes by some desired distance, and then sums them up again to obtain the propagated field. Any set of spatial modes can be used as long as they are complete and orthogonal. For our purposes, we will make use of the LG modes which are given analytically by in Equation \ref{eq:LGdef}. Because the LG modes form a complete and orthogonal basis, we may write any optical field $\ket{U(0)}$ in some initial plane $z = 0$ as follows,}
\begin{equation}
    \ket{U(0)} = \sum_{l,p}  c_{l,p}(0) \ket{l,p,0},
\end{equation}
\CP{\text{where}  $c_{l,p} \equiv \bra{l,p,0}\ket{U(0)}$. If we wish to see how this states evolves in propagation through a vacuum, we must apply the propagation operator $\hat{P}(z) \equiv e^{-i \frac{ z}{2k}\nabla_{\perp}^2}$. As our state is a linear superposition of the basis LG modes $\ket{l,p,0}$, applying the operator onto $ \ket{U(0)}$ applies the operator onto each basis mode as follows,}
\begin{equation}
    \ket{U(z)} = \hat{P}(z)\ket{\Psi(0)} =\sum_{l,p} c_{l,p}(z)\ket{l,p,0} ,    
\end{equation}
 \CP{ where $c_{l,p}(z) = \sum_{l',p'} \bra{0, p ', \ell '} \hat{P}(z) \ket{\ell, p, 0}$, showing that each mode scatters into other modes ($p$-modes for free space) due to propagation as illustrated in Figure \ref{fig:ModalProp} (a). Here we measure the output fields in the same basis as the input fields $\ket{l,p,0}$. Therefore, we stress that the modal coefficients change in propagation and are thus functions of $z$, i.e. $c_{l,p} = c_{l,p} (z)$.}
\CP{Consequently, we see that the modal power spectrum $|c_{l,p}(z)|^2$ is different at the input and output of the channel. However, a one-to-one mapping exists between the input LG modes and their propagated versions, i.e. $\ket{l,p,0}  \rightarrow \ket{l,p,z}$, it follows that if we measure the out put modes in the propagated basis then the coefficients do not change, i.e.,}
\begin{eqnarray}
         \ket{U(z)} = \sum_{l,p} c_{l,p} \ket{l,p,z} 
         \label{eq:preservedbasis}
\end{eqnarray}
where the coefficients
\begin{eqnarray}
         c_{l,p}(z) &=&  \sum_{l',p'} c_{l,p}  \bra{z, p ', \ell '} \hat{P}(z) \ket{\ell, p, 0} \nonumber \\ 
         &=& \sum_{l',p'}  c_{l,p} \langle z, p ', \ell '|\ell, p, z \rangle \nonumber \\  &=&  c_{l,p}
         \label{eq:preservedcoef}
\end{eqnarray}
\CP{are preserved since the output basis remains orthogonal, i.e.  $\langle z, p ', \ell '|\ell, p, z \rangle = \delta_{(\ell, p), (\ell ', p ')}$. We see now that the coefficients are the same in the initial basis as they are in the adjusted basis. This is illustrated visually in Figure \ref{fig:ModalProp} (b), where the modal power spectrum remains unchanged through the channel when the output field is measured in the propagated basis. Going forward, we will always measure the output field using the propagated basis due to the one-to-one mapping between the initial and output basis. This does not necessarily hold for a complex channel with some spatially varying phase perturbation. In this case, we do not only have the propagation operator acting on our input state, but also the phase aberration operator. To see how the mode spectrum is affected, we observe how our two operators affect the input field,}
\begin{eqnarray}
    \hat{T} \ket{U(0)} &=& \hat{P}(L/2) \hat{R}\hat{P}(L/2)\ket{U(0)}, \nonumber \\
     &=& \hat{P}(L/2) \hat{R} \ket{\Psi(L/2)}, \nonumber\\
     &=&\sum_{l,p} c'_{l,p} \ket{l,p,L},
    \label{eq:OutputCoeffs}
\end{eqnarray}
\CP{where sequence of operations include (i) the first propagation operator converting the initial state onto the propagated version ($\ket{U(L/2)}$) expanded in the $\ket{l,p,L/2}$ basis, (ii) the phase operator remapping the coefficients $c_{l_1,p_1} \rightarrow c'_{l,p} = \sum_{l_1,p_1} c_{l_1,p_1} a^{l,p}_{l_1,p_1}$, (iii) which are then preserved after the final propagation operator is applied following Eq. \ref{eq:preservedcoef} and Eq. \ref{eq:preservedbasis} now defined in the transformed basis, $\ket{l,p,L}$. We see here that in general the modal power spectrum will not be preserved due to the presence of the extra coefficient $ a^{l,p_1}_{l,p_2}$ from the phase aberration operator. Thus, to find modes with modal power spectra that are invariant through a complex channel, one must find modes $\ket{U(0)}$ such that $\ket{U(L/2)}$ is an eigenstate of the phase aberration operator $\hat{R}$. In this case we have $\hat{R}\ket{U(L/2)} = e^{i\gamma} \ket{U(L/2)}$, where $e^{i\gamma}$ is some unitary eigenvalue. With this in mind, the computation becomes,}
\begin{eqnarray}
    \hat{T} \ket{U(0)} &=& \hat{P}(L/2) \hat{R}\hat{P}(L/2)\ket{U(0)} \nonumber \\
    &=& \hat{P}(L/2) \hat{R}\ket{U(L/2)} \nonumber \\
    &=& \hat{P}(L/2) e^{i\gamma}\ket{U(L/2)} \nonumber \\
    &=& e^{i\gamma}\ket{\Psi(L)} \\
    &=& e^{i\gamma} \sum_{l,p}  c_{l,p} \ket{l,p,L}    
\end{eqnarray}
\CP{where the coefficients of state are preserved provided the final states are measured in the propagated basis, $\ket{l,p,L}$. With this approach, the modal weights, $c_{l,p}$ are preserved through the channel though the basis elements evolve.}
\CP{The calculation of the operator consists of simple overlap integrals between basis modes and the phase screen, and numerical methods to find the eigenvectors of a matrix are common in many programming languages, including MATLAB and Python. The behaviour of such modes in shown in Figure \ref{fig:ModalProp} (d) where we observe that a mode that is some superposition of basis LG modes that is an eigenstate of the phase aberration operator $\hat{R}$ enters a complex channel and exits with a noticeably different spatial profile, however the modal power spectrum is unchanged through the channel. The modal power spectrum remains unchanged because it is unchanged by the mode-mixing action of the channel but the spatial profile is different because it is reconstructed/measured in a propagated basis. A peculiar quirk of this kind of mode is that its propagation dynamics are the same regardless of the presence of the aberrating medium. The spatial profile of this mode will be the same when passes through its particular complex channel or when it passes through a uniform medium as seen in Figure \ref{fig:ModalProp} (e).  }

\note{In the case of simulating thick/long channels where $\sigma_R^2>1$, one must use multiple phase screens in the simulation. In this case, one should first calculate an operator for each of the phase screens $\hat{R}_i$ with the appropriately propagated basis modes. The total action on the modal spectrum is then the combined action of each phase screen refraction operators, $\hat{R}_{channel} = \hat{R}_N \hat{R}_{N-1}...\hat{R}_2\hat{R}_1$. This channel refractive index operator $\hat{R}_{channel}$ represents the net change in the modal spectrum from the input to the output of the channel. Thus, to apply our approach one needs to then find the eigenvectors of $\hat{R}_{channel}$ to find the invariant spectral modes for a thick/long channel. In both the single phase screen and multiple phase screen cases, one is simply trying to find a particular superposition of input basis modes whose modal spectrum will remain unchanged under the mode mixing action of the channel. As such, we expect the invariant spectral modes in each case to exhibit the same properties, however, the modes for a long/thick medium may require larger basis sizes to help compensate for more complex aberrations and higher modal spreading present in the thicker medium.}

\section{Numerical Simulations}

\begin{figure*}[htbp!]
    \centering
    \includegraphics[width=0.8\linewidth]{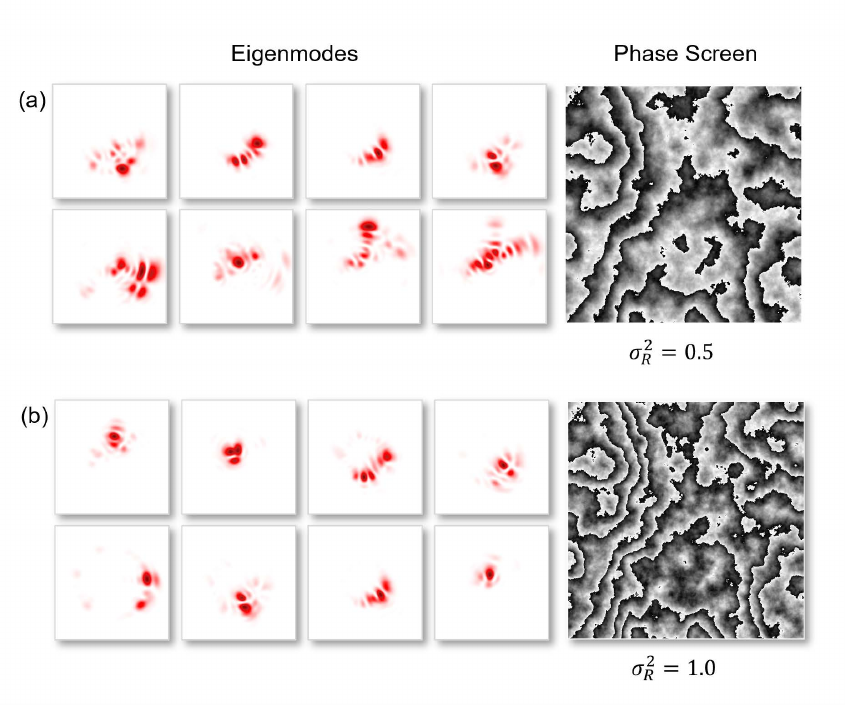} 
    \caption[Invariant spectral modes of turbulence]{\textbf{Invariant spectral modes of turbulence.} The turbulence phase screen and normalised intensity profiles of 8 eigenmodes for two different turbulent channels  with Rytov variance (a) $\sigma^2_R = 0.5$, with Fried parameter $r_0 =9.28$~mm and $D/r_0 = 17.65$ and (b) $\sigma^2_R = 1.0$ with Fried parameter $r_0 =0.61$~mm and $D/r_0 = 26.7$, where $D$ is the transmitting aperture. The channel length in both cases was $100$~m.The operator used to calculate these modes was defined in a basis of 231 LG basis modes with indices ranging from  $l = -10$ to $l = 10$ and $p = 0$ to $p = 10$ and Gaussian beam waist $w_0 = 10$~mm. } 
    \label{fig:ModeCollage}
\end{figure*}

\CP{We performed several numerical simulations to verify our approach. We begun each simulation by defining the parameters of a particular channel. We chose a set length for all channels to be $L=100$~m and a wavelength of $\lambda = 633$~nm. Because we wished to examine channels with only a single phase screen, the maximum Rytov variance of the channel was set to $\sigma_R^2 = 1$. We also used a consistent embedded Gaussian beam waist of $w_0 = 10$~mm for the transmitter basis modes, where $w(z) = w_0 \sqrt{1 + (z/z_R)^2}$ and $z_R = \pi w_0^2/\lambda$ is the Rayleigh range of the beam. We also used a set transmitting aperture diameter of $D=16.4$~cm.}

\CP{As an illustrative example of the spatial profile of our invariant spectral modes, we show 8 of them along with the respective phase screens for two different channels in Figure \ref{fig:ModeCollage}. Figure \ref{fig:ModeCollage} (a) shows the modes for a channel of Rytov variance $\sigma_R^2 = 0.5$ with $D/r_0 = 17.65$ and $r_0 = 9.29$~mm and Figure \ref{fig:ModeCollage} (b) shows the modes for a channel of Rytov variance $\sigma_R^2 = 1.0$ with $D/r_0 = 26.7$ and $r_0 = 6.14$~mm. To calculate these modes, Fourier phase screens where generated using the above mentioned parameters to a resolution of $512\times512$ pixels and 5 subharmonics. The screens were then combined with Equations \ref{eq:ModalPhaseAbbOp} and \ref{eq:ModalPhaseAbbCoeff} to calculate the phase aberration operator $\hat{R}$ for the respective channels. The LG modes that formed the basis had maximum and minimum indices according to: $p_{min} = 0$, $p_{max} = 10$, $l_{min} = -10$ and $l_{max} = 10$. The total basis size was then 231 basis modes. In-built MATLAB functions were then used to determine the eigenstates of the phase aberration operator.} Upon visual inspection, the invariant spectral modes have quite a complex spatial structure with both high intensity and low intensity regions and lack any obvious symmetries. This is in contrast to the symmetries seen in equivalent modes in free space, namely Laguerre-Gaussian (LG) and Hermite-Gaussian (HG) beams which have polar and Cartesian symmetries, respectively. This is due to the fact that the channels, and specifically the phase screens, have no obvious symmetries. Upon closer inspection, we observe that the invariant spectral modes in both cases have features that resemble those of HG beams and may even look like distorted or perturbed HG modes. We can possibly understand this resemblance by referring back to the analogies drawn in between Equations \ref{eq:paraxial helmholtz} and \ref{eq:2DTDSE}. In the traditional approaches of quantum mechanical perturbation theory, one may find the eigenstates of a perturbed system with a perturbed Hamiltonian by describing them as superpositions of eigenstates of the unperturbed Hamiltonian if the perturbation is small. This typically leads to the eigenstates of the perturbed system looking like perturbed eigenstates of the unperturbed system. This intuition carries over to our turbulence (perturbed) channel, whose invariant spectral modes look like perturbed eigenmodes of a free space (unperturbed) channel.

\subsection{Robust Modal Spectra}

\begin{figure*}[htbp!]
    \centering
    \includegraphics[width=\linewidth]{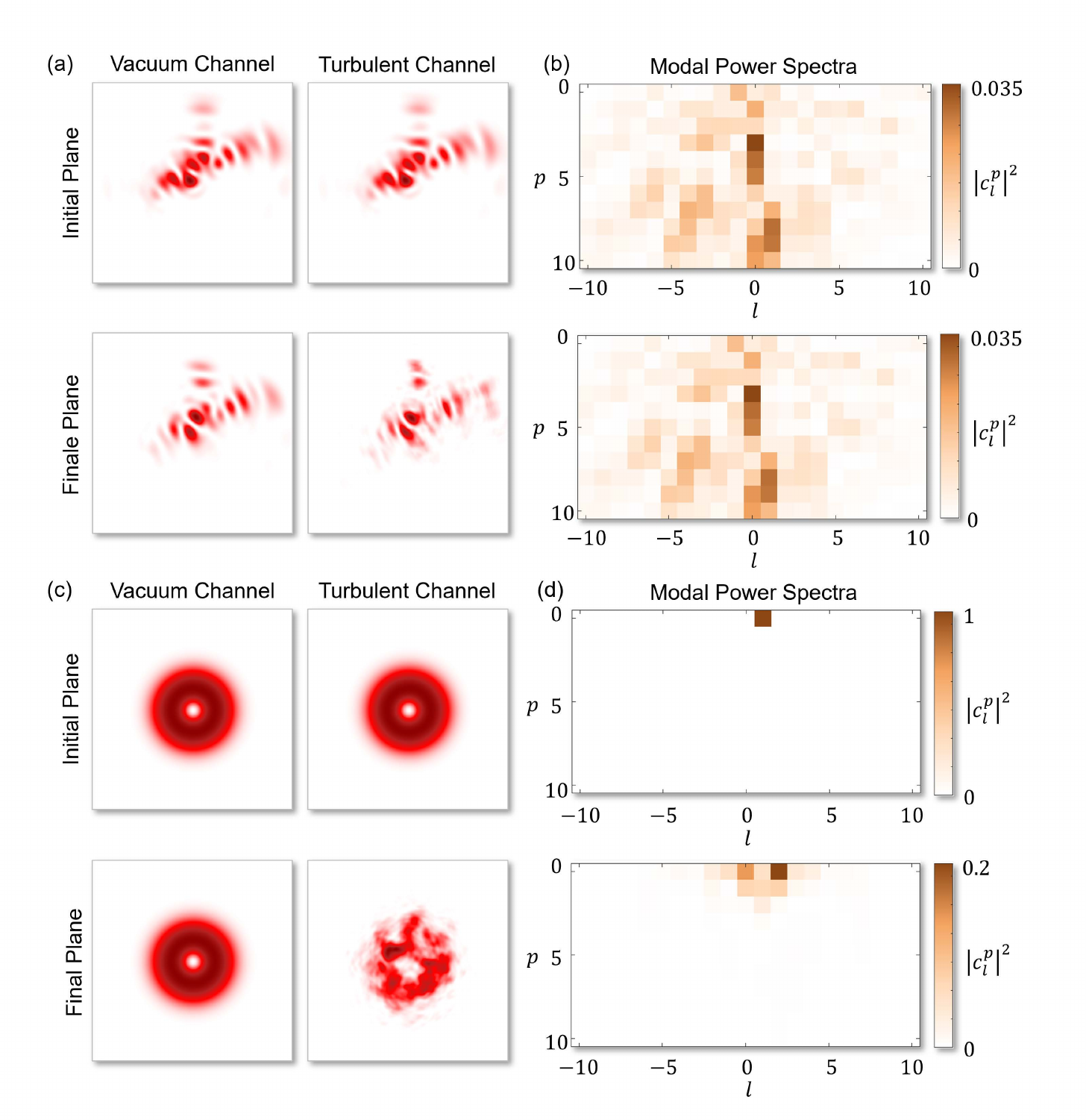} 
    \caption[Spectrum robustness in a 231 dimensional basis]{\textbf{Spectrum robustness in a 231 dimensional basis.} (a) The intensity profiles of an eigenmode calculated with an operator with 231 basis modes before and after propagating through a vacuum channel (left panels) and a turbulent channel (right panels). (b)  The modal power spectrum $|c_p^l|^2$ of the eigenmode before (top) and after (bottom) propagating through the turbulent channel. (c) The intensity profiles of an LG mode with $p=0$ and $l=1$ after propagating through a vacuum channel (left panels) and through the same turbulent channel (right panels).  (d) The modal power spectrum $|c_p^l|^2$ of the LG before (top) and after (bottom) propagating through the turbulent channel.} 
    \label{fig:SpectraRobustness231}
\end{figure*}

\begin{figure*}[htbp!]
    \centering
    \includegraphics[width=\linewidth]{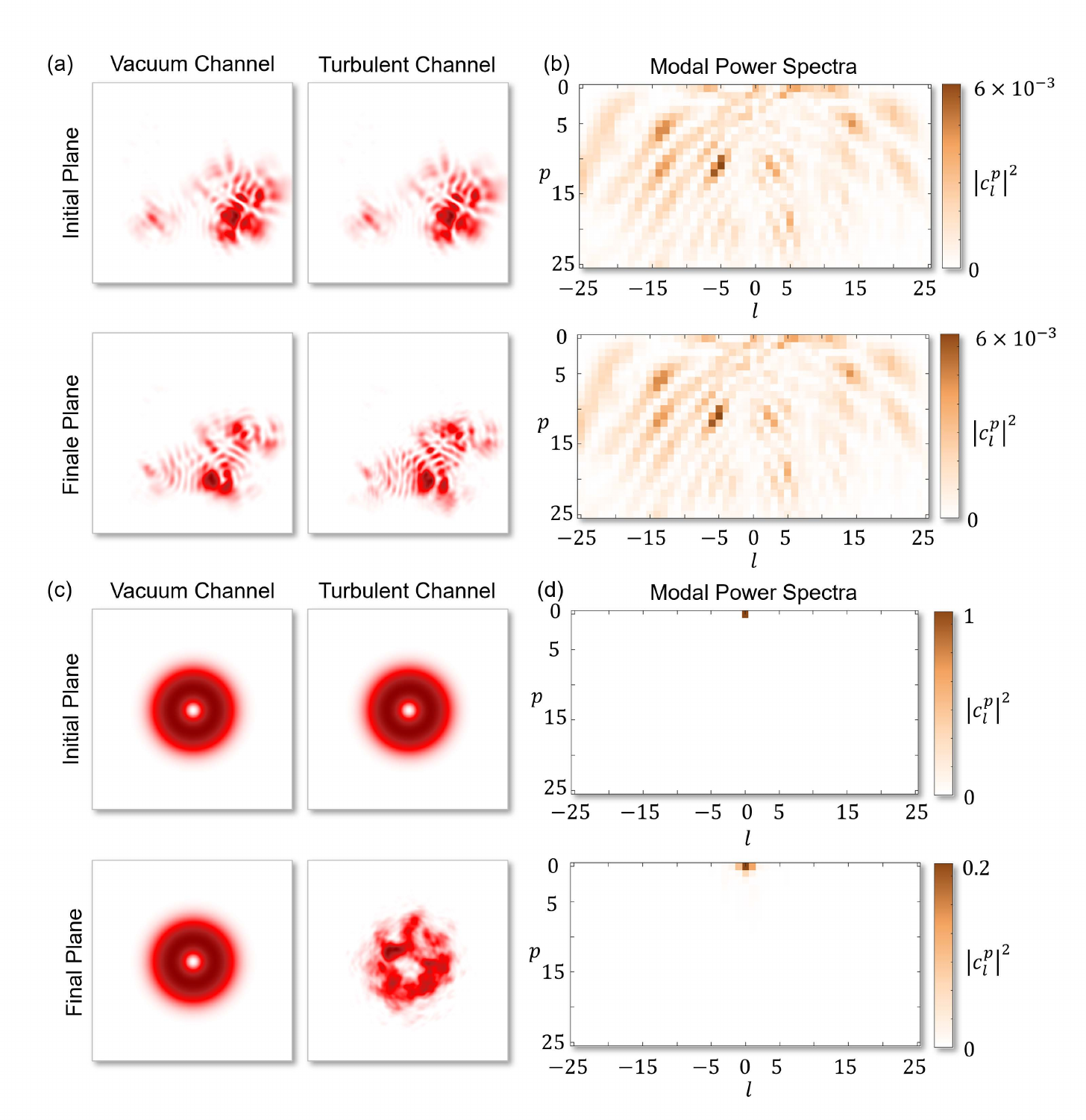} 
    \caption[Spectrum robustness in a 1326 dimensional basis]{\textbf{Spectrum robustness in a 1326 dimensional basis.} (a) The intensity profiles of an eigenmode calculated with an operator with 1326 basis modes before and after propagating through a vacuum channel (left panels) and a turbulent channel (right panels). (b)  The modal power spectrum $|c_p^l|^2$ of the eigenmode before (top) and after (bottom) propagating through the turbulent channel. (c) The intensity profiles of an LG mode with $p=0$ and $l=1$ after propagating through a vacuum channel (left panels) and through the same turbulent channel (right panels).  (d) The modal power spectrum $|c_p^l|^2$ of the LG before (top) and after (bottom) propagating through the turbulent channel.} 
    \label{fig:SpectraRobustness1326}
\end{figure*}

In order to test the robustness of the calculated invariant spectral modes, their propagation through their respective channels was simulated numerically using \CP{angular spectrum propagation. Angular spectrum propagation was chosen as it is independent of the modal propagation approach used to identify the invariant spectral modes, andis also an alternative to the Fresnel propagator used to compute the pixel eigenmodes later on in this work. The channel was simulated according to a single phase screen simulation as described in Ref \cite{peters2025structured} and as illustrated in Figure \ref{fig:Operator} (a). The mode was propagated halfway through the channel using a numerically implementation of the angular spectrum procedure. The calculated turbulent phase screen was then added to the beam. The beam was then propagated the remaining half of the channel, again using the angular spectrum method.} An example \CP{of the invariant modal spectrum} is shown in Figure \ref{fig:SpectraRobustness231} (a), where we show a mode for a channel with Rytov variance $\sigma_R^2 = 0.5$ and all the same parameters as mentioned previously. The top left and bottom left panels show the beam before and after propagating through a uniform/vacuum channel with no aberration. The top right and bottom right panels show the beam before and after propagating through the aberrated channel. We see that the mode's intensity profile before and after propagating through the channel in both cases is different. However, when we examine the modes' modal power spectrum, i.e., the coefficients $|c_l^p|^2$, before and after the channel as shown in Figure \ref{fig:SpectraRobustness231} (b), we see that it has remained almost perfectly unchanged. This example illustrates what we expect: the modal power spectrum is invariant to the action of the channel while the mode's spatial profile is dependent on the amplitude and phase profile of the basis modes at that plane where it is being measured. Because the modal power spectrum of the invariant spectral mode is unchanged by the channel, and because any free space propagation also does not change the modal spectrum of any complex field, we expect the the mode's spatial profile in the final plane for both the vacuum channel and the turbulent channel to be the same. While both profiles at the final plane are similar in structure, the invariant spectral mode at the output of the turbulent channel exhibits noticeable high-frequency irradiance variations. This is possibly due to size of the spatial mode basis. The basis set of 231 modes is perhaps too small to capture some of the high-frequency spatial variations and thus cannot compensate for them when constructing the eigenmode. This can be easily fixed by simply increasing the size of the spatial mode basis used to construct the channel operator. However, we still see that the modal spectrum is unchanged within the 231 LG modes that make up our modal basis. Figure \ref{fig:SpectraRobustness1326} shows the same results as Figure \ref{fig:SpectraRobustness231} but for a spatial mode basis size of 1326 LG modes with indices ranging from $l = -25$ to $l = 25$ and $p = 0$ to $p = 25$ and Gaussian beam waist $w_0 = 10$~mm. Immediately, we see that the invariant spectral mode before propagating through the channel has very fine and high-frequency features when compared to the invariant spectral mode in Figure \ref{fig:SpectraRobustness231} (a).

\begin{figure*}[htbp!]
    \centering
    \includegraphics[width=\linewidth]{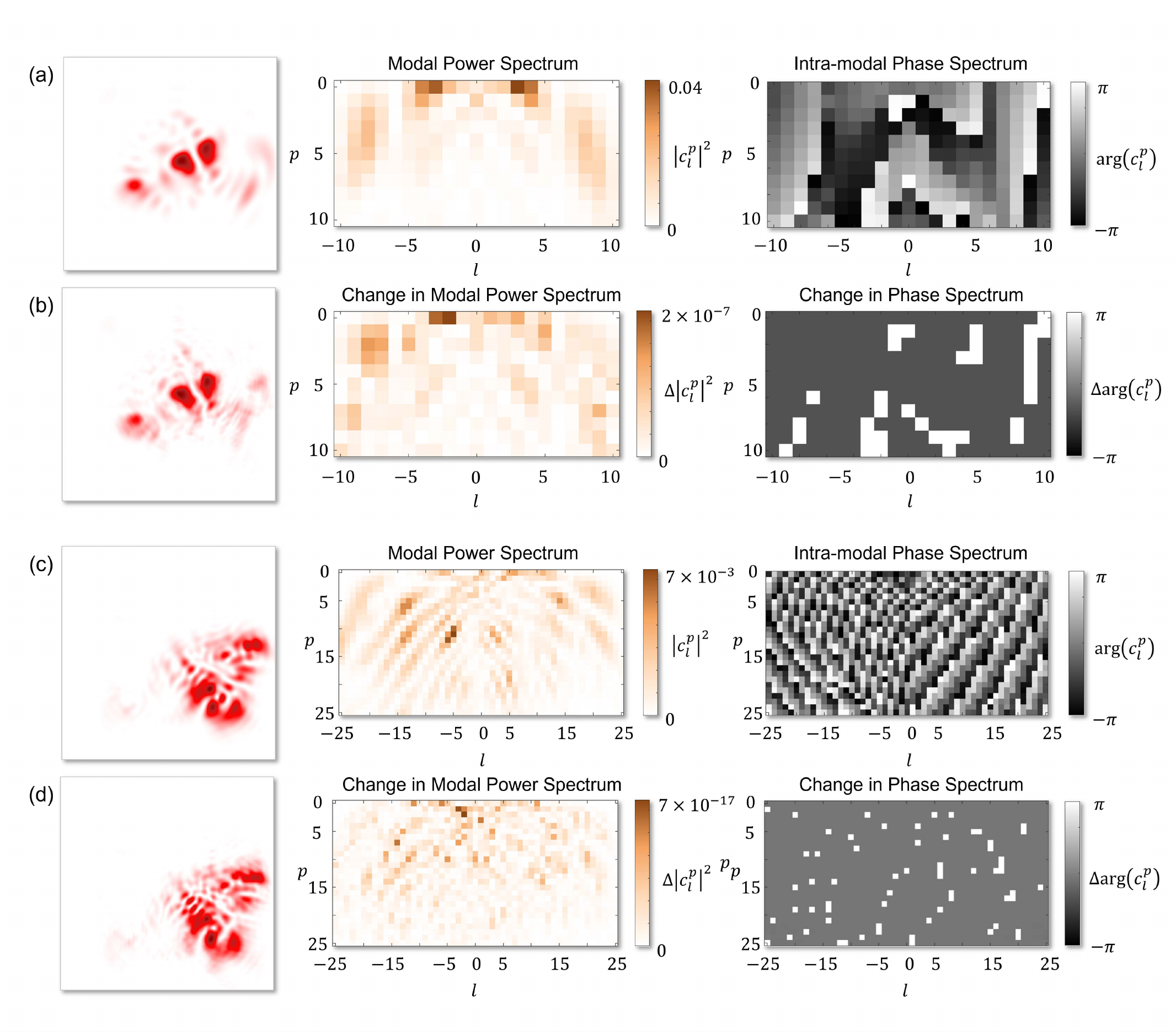} 
    \caption[Changes in the modal spectra]{\textbf{Changes in the modal spectra.} (a) The intensity profile, modal power spectrum  $|c_p^l|^2$ and intramodal phase spectrum $\text{arg}(c_p^l|^2)$ of an invariant spectral mode calculated for a channel of Rytov variance $\sigma^2_R = 0.5$ with Fried parameter $r_0 =9.28$~mm and $D/r_0 = 17.65$ and channel length of $100$~m before propagating through the channel. The basis consisted of 231 basis modes. (b) The intensity profile of the eigenmode in (a) after propagating through the turbulent channel, the change in its modal power spectrum  $\Delta |c_p^l|^2$ and the change in its intramodal phase spectrum $\Delta \text{arg}(c_p^l|^2)$. (c) The intensity profile, modal power spectrum  $|c_p^l|^2$ and intramodal phase spectrum $\text{arg}(c_p^l)$ of an invariant spectral mode calculated for a channel of Rytov variance $\sigma^2_R = 0.5$ with Fried parameter $r_0 =9.28$~mm and $D/r_0 = 17.65$ and channel length of $100$~m before propagating through the channel. The basis consisted of 1326 basis modes. (d) The intensity profile of the eigenmode in (c) after propagating through the turbulent channel, the change in its modal power spectrum  $\Delta |c_p^l|^2$ and the change in its intramodal phase spectrum $\Delta \text{arg}(c_p^l|^2)$.} 
    \label{fig:SpectrumChanges}
\end{figure*}

We now turn to a more explicit example of how the modal spectrum of the invariant spectral mode is affected by the channel. Figure \ref{fig:SpectrumChanges} (a) and (b) shows the intensity, modal power spectrum $|c_p^l|^2$, intramodal phase spectrum $arg(c_p^l)$ and the change in these two spectrums for an invariant spectral mode of a channel of Rytov variance $\sigma^2_R = 0.5$ with Fried parameter $r_0 =9.28$~mm and $D/r_0 = 17.65$ and channel length of $100$~m. The basis used to calculate them consisted of 231 LG modes with indices ranging from $l = -10$ to $l = 10$ and $p = 0$ to $p = 10$ and Gaussian beam waist $w_0 = 10$~mm. The intensity of the mode shows the same trends that we have observed previously. We immediately see that both the modal power spectrum and the intramodal phase spectrum both contain noticeable structure, where basis LG modes with higher $l$-indices are more likely to also have larger $p$-indices. Most importantly however, we can see that the change in the modal power spectrum is 5 orders of magnitudes smaller than the modal powers themselves, indicating that this spectrum remains virtually unchanged during propagation through the channel. On the other hand, the intramodal phase spectrum does noticeably change in propagation. Thus, we can conclude that any loss in fidelity in the eigenmodes is solely due to changes in the intramodal phases in the LG basis modes that make up the eigenmode. Interestingly, most of the LG basis modes maintain the same intramodal phase, and those that do change always gain a phase of approximately $\pi$. Figures \ref{fig:SpectrumChanges} (c) and (d) show the same results for an eigenmode of the same channel and \CP{same realization of turbulence}, except that the LG basis used consisted of 1326 basis modes. We see similar features to the modal power spectrum of eigenmode from the smaller basis, but see that the change in the modal power spectrum is significantly smaller. This indicates that the increased number of basis modes in the calculation of the channel operator results in more robust eigenmodes. However, given that the change in the modal power spectrum for the eigenmodes the 231 modal basis is very small, this on its own might not be a significant enough advantage to justify the increased measurement and computation time required to implement the larger basis size. The better performance of the larger basis modes can be seen when we look at the intramodal phase spectrums. While we do see again that some of the LG basis modes see no change in intramodal phase while others see a $\pi$ phase shift, the proportion of modes that see this phase shift is about half as many as compared to the eigenmodes in the 231 sized basis. This leads to a noticeable improvement in the fidelity of the eigenmodes at the output of the channel. We therefore see that for both basis sizes, the modal power spectrum is perfectly robust through the turbulent channel. Meaning that the power in each component basis modes does not change in propagation through the channel resulting in invariant spectral modes that remain remarkably robust through an highly aberrating medium. We also see that a larger basis size increases the invariant spectral mode' ability to maintain the intramodal phases in the superposition to generate a particular mode. However, invariant spectral mode calculated in the smaller basis are still able to maintain a high degree of fidelity through the channel.

\subsection{Average Mode Fidelity and Orthogonality}

\begin{figure*}[h!]
    \centering
    \includegraphics[width=\linewidth]{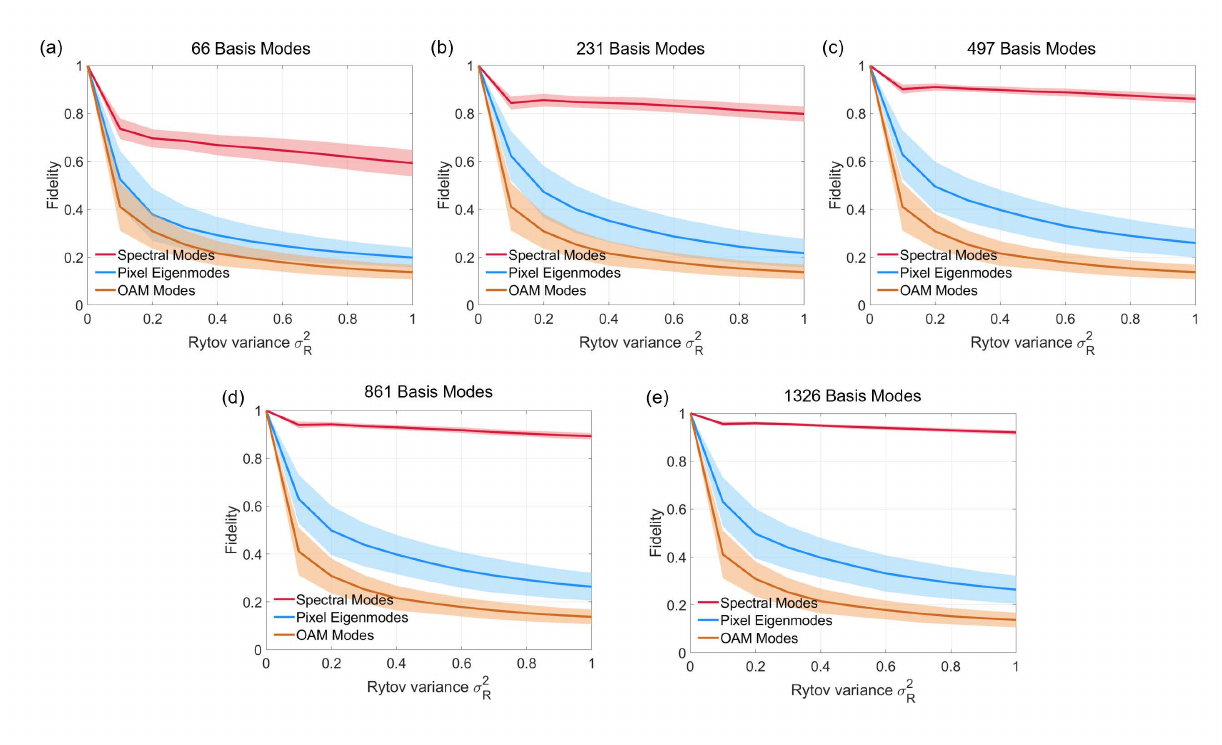} 
    \caption[Fidelity through complex channels]{\textbf{Fidelity through complex channels.} The average fidelities of the invariant spectral modes (red), pixel eigenmodes (blue) and OAM modes (orange) with basis sizes of (a) 66 LG modes and 81 pixel modes, (b) 231 LG modes and 256 pixel modes, (c) 497 LG modes and 529 pixel modes, (d) 861 LG modes and 900 pixel modes and (e) 1326 LG modes 1600 pixel modes. Each data point was averaged over 100 turbulence realizations, taking the average fidelity of the first 20 modes for each realization. } 
    \label{fig:FidelityPlot}
\end{figure*}

\CP{So far we have only looked at specific examples of invariant spectral modes which can maintain their modal power spectrum through simulated complex channels. To investigate the average behaviour of these modes, we performed larger scale simulations as seen in Figures \ref{fig:FidelityPlot}. Each plot shows the average fidelity of 20 invariant spectral modes computed for 100 different channel realizations for each turbulent strength. We compare this to the performance of traditional OAM mode and pixel eigenmodes calculated in a comparable sized basis. The basis size are Figure \ref{fig:FidelityPlot} (a) 66 LG modes and 81 pixel modes, (b) 231 LG modes and 256 pixel modes, (c) 497 LG modes and 529 pixel modes, (d) 861 LG modes and 900 pixel modes and (e) 1326 LG modes 1600 pixel modes. We tested channels of turbulence strengths ranging from $\sigma_R^2 = 0$ to $\sigma_R^2 = 1.0$ in steps of $0.1$. The distortions strengths for these channels were $D/r_0 = 0, 6.75, 10.2, 13, 15.5, 17.65, 19.7, 21.5, 23.44, 25.26,$ and $ 26.7$, with all other parameters the same as previously described. The performance of the invariant spectral modes is shown in red, the pixel eigenmodes in blue and the OAM modes in orange. The shaded region indicates the standard deviations. The fidelity of the output mode was calculated by taking its inner product with the predicted mode based on the respective method being employed. We see that for every basis size employed, the invariant spectral modes perform best, with fidelities between 40\%-70\% higher, depending on the basis size and turbulence strength. The next best are the pixel eigenmodes, which perform closer to the OAM modes but do show some improvement over them. This is expected as they are calculated specifically to compensate for the channel's aberration. Their poor performance can be attributed to the fact that the number of pixels being used is not sufficient to capture and compensate for the perturbations in the channel. The OAM modes perform the worst out of the three. This again is to be expected as nothing is being done by the modes to compensate for the channel's distortion. Additionally, we see that for all three approaches, the fidelity of the output modes decreases as the strength of the channel's perturbation $\sigma_R^2$ increases. This again is to be expected as a larger channel perturbation will have higher frequencies in the phase screen and thus a more complex structure to compensate for. However, the performance of the invariant spectral modes deteriorates much more slowly than the performance of the pixels eigenmodes and the OAM modes, indicating our approach holds up significantly better at stronger and stronger distortions. We also see that the performance of the invariant spectral modes is a lot more consistent for each turbulent strength. This is clearly seen in the fact that the standard deviation of the invariant spectral modes is comparable to the other methods in Figure \ref{fig:FidelityPlot} (a) but noticeably smaller in Figures (b)-(e) over all of the turbulence strengths. The standard deviation of the invariant spectral modes also decreases as the basis size increases to a point where it is almost unnoticeable at a basis size of 1326 modes. This adds further evidence to support that the larger basis size is better at completely describing the action of the channel and thus is able to produce modes that are robust to its action. }

\begin{figure*}[h!]
    \centering
    \includegraphics[width=\linewidth]{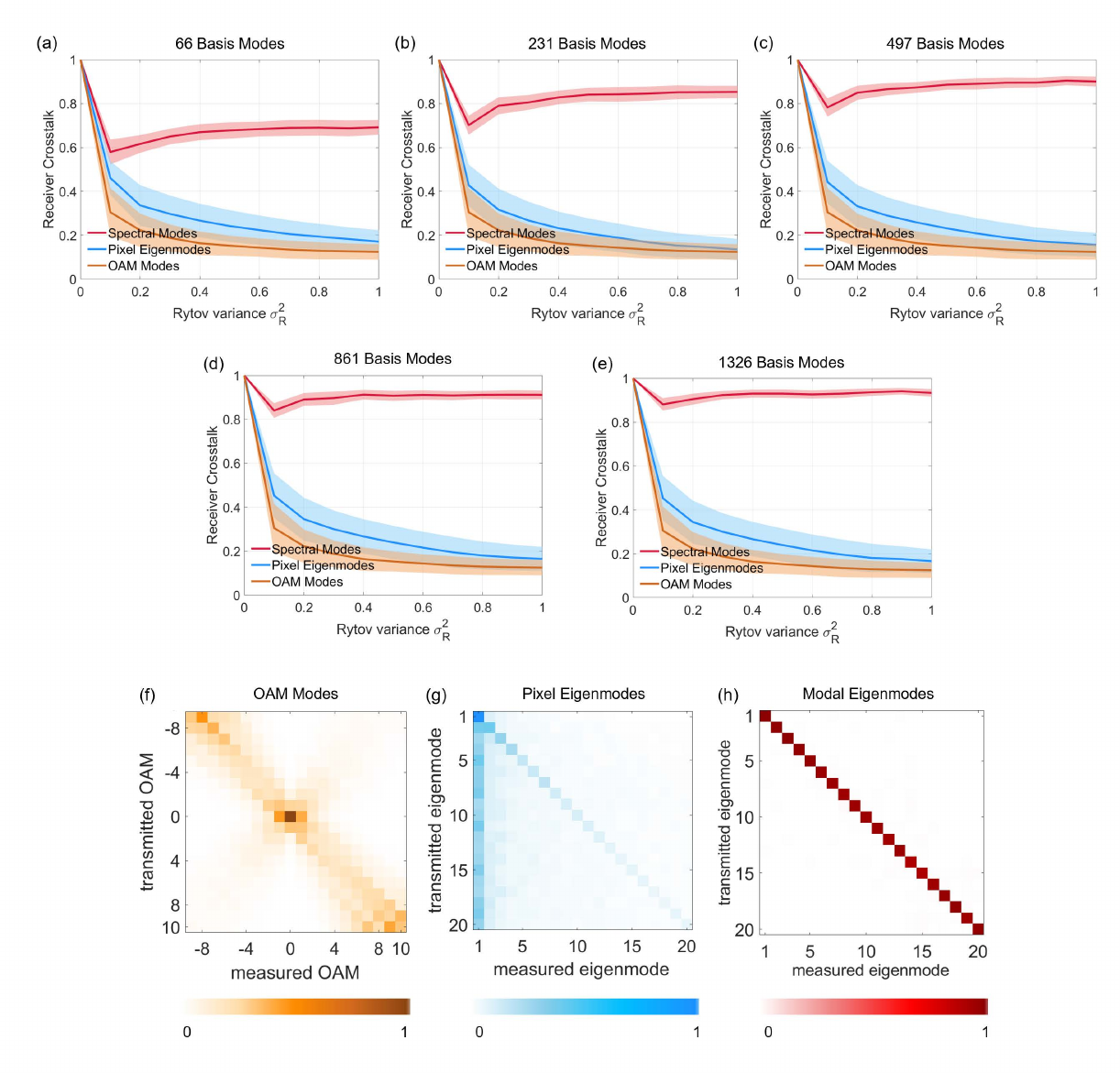} 
    \caption[Orthogonality through complex channels]{\textbf{Orthogonality through complex channels.} The average Orthogonality of the invariant spectral modes (red), pixel eigenmodes (blue) and OAM modes (orange) with basis sizes of (a) 66 LG modes and 81 pixel modes, (b) 231 LG modes and 256 pixel modes, (c) 497 LG modes and 529 pixel modes, (d) 861 LG modes and 900 pixel modes and (e) 1326 LG modes 1600 pixel modes. Each data point was averaged over 100 turbulence realizations, taking the average orthogonality of the first 20 modes for each realization. We also show the averaged crosstalk matrices for the (f) OAM modes, (g) pixel eigenmodes and (h) pixel eigenmodes for channels of strength $\sigma_R^2 = 1.0$ with basis sizes of 1326 LG modes 1600 pixel modes.}  
    \label{fig:OrthogonalityPlot}
\end{figure*}

Another important property of spatial modes for use in applications such as imaging and communications is that of orthogonality. Orthogonality allows one to easily distinguish and separate out the contributions of different spatial modes. It also reduces the number of measurements needed to accurately reconstruct an image. Because we are interested in using the modes for energy and information transport through a complex medium, the important quantity to look at is the crosstalk at the receiver to see if the orthogonality of the modes in the transmitted basis is maintained through the channel. To do this we compare the average fidelity of the crosstalk matrices for the three different mode sets at the receiver in Figures \ref{fig:OrthogonalityPlot} (a)-(e), where $1$ represents perfect fidelity, minimal crosstalk and thus ideal orthogonality at the receiver while $0$ represents maximal crosstalk and complete loss of orthogonality at the receiver. \CP{We examine the orthogonality of the modes for 5 different basis sizes for both the invariant spectral modes and pixel eigenmodes. The parameters for these modes are the same as seen in Figure \ref{fig:FidelityPlot} (a)-(e). We compare each of these cases to the crosstalk observed by the OAM modes after propagating through the channel. We observe that for all turbulence strengths and for all basis sizes, the receiver crosstalk is significantly better for the invariant spectral modes as compared to the pixel eigenmodes and the OAM modes by ~15\%-40\% at low turbulence strengths to 70\% at higher turbulence strengths. The pixel eigenmodes perform marginally better than the OAM modes by 5\%-10\%. For the pixel eigenmodes and the OAM modes, we see that the receiver crosstalk decreases as the turbulence strength increases. This is expected as higher turbulence strengths will have a stronger perturbative effects on the beam and scrambling the modal contents in the original basis. In an ideal case, the pixel eigenmodes would exhibit a higher degree of orthogonality even at these higher strengths \cite{klug2023robust}. However, Figure \ref{fig:FidelityPlot} shows that the fidelity of the eigenmodes is very low for these channels and so we would not necessarily expect them to maintain their orthogonality through the channel.We see that for the invariant spectral modes, the receiver crosstalk is consistently above 80\% for the larger basis sizes and shows a slight increase as the turbulence strength increases for all basis sizes. Figure \ref{fig:OrthogonalityPlot} (f),(g) and (h) show the averaged crosstalk matrices for OAM modes, pixel eigenmodes and invariant spectral modes respectively. These all use the same parameters as previously described for a channel of Rytov variance $\sigma_R^2 = 1$. The OAM mode crosstalk matrix shows spreading of the transmitted modes into neighbouring modes, which is expected from the intramodal coupling behaviour experienced by the complex channel. The crosstalk matrix for the pixel eigenmodes exhibits two distinct features. One is a noticeable diagonal line with very little intermodal coupling around this line. This shows some correspondence between the transmitted and received mode due to the construction of the modes from the channel operator. The other feature is strong coupling of each eigenmode to the lowest order eigenmode. Due to the modes not maintaining a high fidelity at the output of the channel, we can attribute this feature to the inability of the eigenmodes to maintain their structure through the channel. The crosstalk matrix for the invariant spectral modes shows an almost perfect one-to-one correspondence indicating a high degree of orthogonality, even at the output of the channel.   }

\subsection{Communicating with Eigenmodes}

\begin{figure*}[t!]
    \centering
    \includegraphics[width=\linewidth]{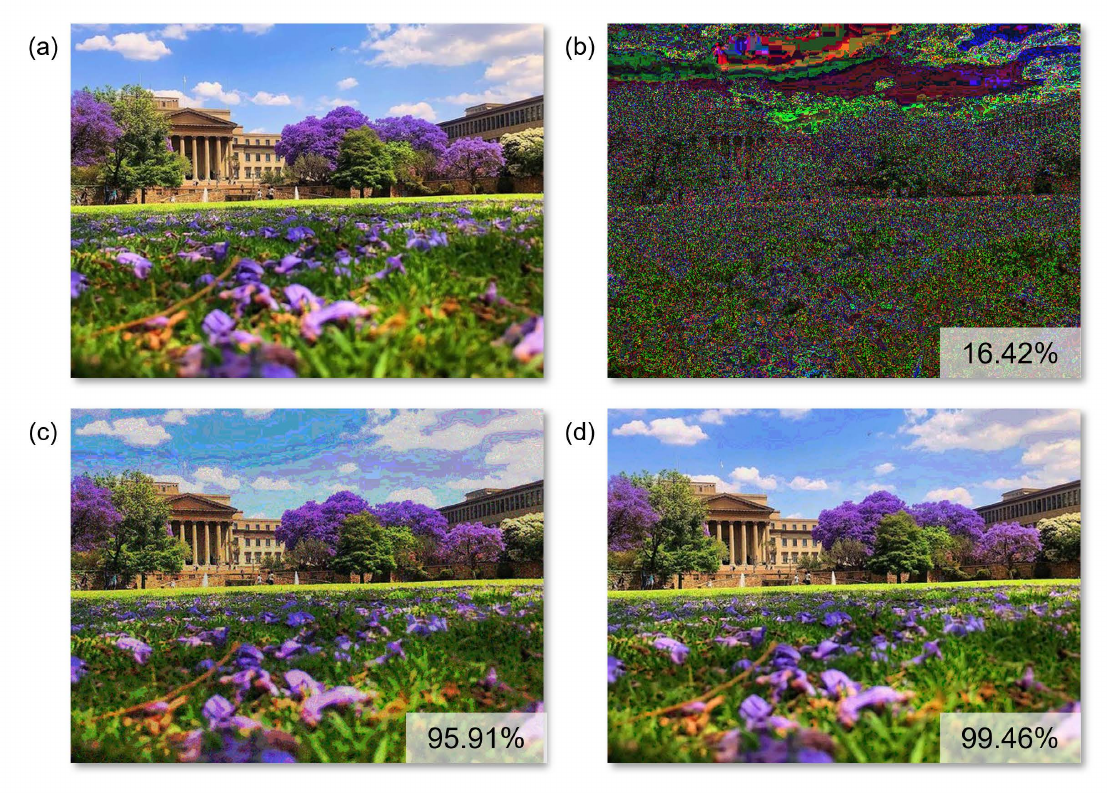} 
    \caption[Robust communication through complex channels]{\textbf{Robust communication through complex channels.} (a) The original 8-bit image with a resolution of $423\times564$. (b) Simulated reconstruction of the image using OAM encoding through a turbulent channel of $\sigma_R^2 = 0.5$. Eight OAM modes were used from $l = -3$ to $l = 4$. (c) The simulated reconstruction of the image through the same channel using eigenmodes to encode the information using a channel operator with 231 basis modes. (d) The simulated reconstruction of the image through the same channel using eigenmodes to encode the information using a channel operator with 1326 basis modes.   } 
    \label{fig:Comms}
\end{figure*}

One of the benefits of knowing the eigenmodes of an aberrated channel is that they can be used to transport information through the channel without it being distorted or lost. To demonstrate \CP{how the invariant spectral modes derived in this paper can be used in the same manner,} we simulate the transmission of an 8-bit RGB image with a resolution of $423\times564$ through a static, aberrated channel with a turbulent phase screen of Rytov variance $\sigma_R^2 = 0.5$ and length of $100$~m. Figure \ref{fig:Comms} (a) shows the image to be transmitted. We used three different sets of 8 modes to transmit the image information through the channel and reconstruct it at the receiver. These included 8 OAM modes with $l = -3$ to $l = 4$ shown in Figure \ref{fig:Comms} (b), modal eigenmodes calculated from an operator with 231 basis modes shown in Figure \ref{fig:Comms} (c), and modal eigenmodes calculated from an operator with 1326 basis modes Figure \ref{fig:Comms} (d). Our encoding approach involves associating each of the 8 modes in a set with one of the digits in the 8-bit string. Thus, for each 8-bit string representing the greyscale value of a pixel in a particular colour channel, a superposition of modes is sent. Modes absent from the superposition correspond to a 0 digit, while modes present in the superposition correspond to a one digit. The 8-bit string at the receiver can then be easily determined by a modal decomposition of the transmitted superposition into the basis being used. As expected, the use of OAM modes results in an image reconstruction that is severely distorted, with a correlation of $16.42\%$ to the original image. In contrast, the use of invariant spectral modes with a basis size of 231 modes shows a significant improvement in maintaining the fidelity of the image with a correlation of $95.91\%$ to the original image. Increasing the basis size of the channel operator leads to even more improvement, where eigenmodes calculated with an operator with 1326 modes result in the transmitted image having a $99.46\%$ correlation to the original image.

\section{Practical Implementation}

\CP{So far, we have demonstrated that the invariant spectral modes proposed in this work exhibit a high degree of robustness through channels that would otherwise highly aberrate other forms of structured light modes. We have also shown that these modes maintain a high degree of orthogonality even after propagating through the complex channel. Both of these properties indicate that these modes can be used to transport information or image an object through a complex channel. We have gone through some theoretical and numerical details to verify that this can be done, but this detail does not directly relate to the experimental implementation of this approach. This detail is necessary to ensure that our numerical simulation is accurately replicating a channel but in experiment, most of this work will be done by nature. Thus, in experiment, all that is needed is to obtain the transmission matrix of a channel. This is traditionally done by probing the channel one input basis mode at a time and measuring which output basis mode it maps to  \cite{popoff2010measuring, popoff2011controlling}. Once the transmission matrix is obtained, one may use Equation \ref{eq:OutputCoeffs} to determine the phase aberration operator $\hat{R}$ and then calculate the invariant spectral modes of the channel. Our approach makes this process much more efficient as the size of the input and output bases can be much smaller than previous approaches that used the pixel basis (231 invariant spectral modes vs 1600 pixel modes) while still maintaining a high dimensional basis set with which to image or encode information. A potential challenge is the on-demand generation of the basis modes for probing the channel and the subsequent generation of the invariant spectral modes. This can be easily overcome with the use of digital devices such as liquid crystal spatial light modulators (SLMs), digital micro-mirror devices (DMDs) and the recently developed phase-only light modulators (PLMs) \cite{rocha2024fast} which can all perform high fidelity complex amplitude beam shaping for arbitrary beam types at very high speeds. A much greater challenge is that of measuring modes at the output of the channel and it is here where our approach has a significant advantage over previous attempts. By greatly reducing the number of modes needed to reconstruct the channel, the number of measurements needed at the output is significantly reduced. This makes performing a channel tomography using the principle of modal decomposition and projective measurements with SLMs, DMDs or PLMs quite feasible. One can also make use of multi-plane light converters (MPLCs) or other static projective optics, which have demonstrated the ability to accurately sort large numbers of spatial modes \cite{fontaine2019laguerre}.}

\CP{We have made use of atmospheric turbulence as an example as it dynamics and effect on structured light beams are very well understood as compared to other media, however, the approach outlined here is valid for other media as well, including biological tissues, optical fibres, underwater turbulence and thermally aberrated systems. Many of these channels are not static, and do evolve in time and therefore changing the set of invariant spectral modes that will successfully propagate through the channel. In the case of these media, one must have a way of updating the channel's transmission matrix to keep on applying this technique. Depending on the medium, this can be quite challenge. Slowly time-varying media such as biological tissue or optical fibres change on timescales (several seconds to minutes) in which it would be easy to perform a full measurement an determine the appropriate set of invariant spectral modes. However, more extreme cases such as atmospheric and oceanic turbulence change at timescales on the order of milliseconds. In these extreme cases, the implementation of our approach is possible, especially given the speeds at which devices such as DMDs, photodiodes, and MPLCs can operate (which is on the order of kHz), but does present a significant engineering challenge that should not be downplayed. It is here again where our approach does show improvements over previous attempts, where the reduced basis size require much shorter measurement times making the implementation of high dimensional structured light bases far more likely. Such challenges are also well suited to predictive algorithms \cite{valzania2023online} and machine learning approaches which can possibly predict the evolution of the channel.  }

\section{Conclusion}

Finding ways of transporting light through aberrated media in an invariant manner remains a pressing challenge with the promise of greater information transfer speeds, improved precision in measurements, and high resolution biological imaging. In this work, we have demonstrated a procedure for using well-known descriptions of complex media and defining them in a new basis built on spatial modes of light. We then used this procedure to describe the action of a highly turbulent channel in LG basis. This operator was subsequently used to find the invariant modes of the channel and their invariant nature was confirmed through numerical simulations. \CP{The modes we found leverage the one-to-one mapping between LG beams and their scaled versions upon propagation, allowing us to find invariant modes that preserve their modal spectrum even though their relative phases may change}. These invariant spectral modes show high fidelities while using substantially fewer basis modes than previous attempts. Our results emphasize the importance of choosing a basis that best suites the system one is trying to measure or probe as it will greatly simplify the medium's description. While we have used OAM and turbulence as our example problem, the formulation of our approach is neither channel nor modal basis dependent. It can easily be extended to many forms of complex media, including optical fibres, underwater channels and turbid systems.

\section*{Disclosures}
The authors declare no conflicts of interest.

\section*{Data Availability}
Data underlying the results presented in this paper are not publicly available at this time but may be obtained from the authors upon reasonable request.

\end{document}